\documentclass[a4paper]{jpconf}
\usepackage{graphicx}
\begin{document}
\title{ The Nuclear Shell Model Toward the Drip Lines}
\author{ A. Poves$^*$, E. Caurier$^{\dagger}$,  F. Nowacki$^{\dagger}$ and K. Sieja$^{\dagger}$}

\address{$^*$Departamento de F\'{\i}sica Te\'orica
  and IFT/UAM-CSIC, Universidad Aut\'onoma  de Madrid, 28049-Madrid, Spain\\
  $^{\dagger}$IPHC, IN2P3-CNRS/Universit\'e Louis Pasteur, 67037-Strasbourg France.}

\ead{alfredo.poves@uam.es}

\begin{abstract} 

We describe the "islands of inversion" that occur  when approaching the neutron drip line  around the
magic numbers N=20, N=28 and N=40 in the framework of the Interacting Shell Model in very large valence 
spaces. We explain these configuration inversions (and the associated shape transitions) as the result of the
competition between the spherical mean field (monopole) which favors magicity and the correlations (multipole)
which favor deformed intruder states. We also show that the N=20 and N=28  islands are in reallity a single
one,  which for the Magnesium isotopes is limited by  N=18 and  N=32.

\end{abstract}

\section{Monopole anomalies and Multipole universality}
\label{intro}
The different facets of the nuclear dynamics depend on the balance of the two main components of the nuclear hamiltonian;
  the Monopole which produces the effective spherical mean field and the Multipole responsible for the correlations \cite{Duf96}.
  Large scale shell model calculations have unveiled the monopole anomalies
of the two-body realistic interactions, $i.e$ that they tend to produce effective single 
particle energies which are not compatible with the experimental data and which, if used 
without modifications, produce spectroscopic catastrophes.
Already in the late 70's Pasquini and Zuker \cite{Pas76} showed that the Kuo Brown  \cite{Kuo68}
interaction could not produce neither a magic  $^{48}$Ca nor a magic  $^{56}$Ni. In this last
case it  made a nearly perfect rotor instead.  A few monopole corrections (mainly T=1)
restored high quality spectroscopy.    Otsuka {\it et al.} \cite{Ots10} have recently  shown that the monopole component of the three body force
  may explain the monopole anomalies relevant for  $^{28}$O and  $^{48}$Ca.
  The Multipole component of the realistic two body interactions
(dominated by L=0 pairings, quadrupole and octupole) does not seem to require any substantial modification 
and it is "universal" in the sense that all the
interactions produce equivalent multipole hamiltonians.
Magic numbers are associated to energy gaps in the spherical mean
field. Therefore, to promote particles above the Fermi level
costs energy. However, in some cases  intruder
configurations can compensate  their loss of monopole energy  with
their huge gain in correlation energy.
Several  examples
of this phenomenon exist in stable magic
nuclei in the form of coexisting
spherical, deformed and superdeformed states in a very narrow energy range, providing
examples of  nuclear allotropy.
In the case of $^{40}$Ca they can be described in the spherical shell model framework  \cite{Cau07}.

\section{The islands of inversion at N=20 and N=28 far from stability}
\label{sec:1}

The region around $^{31}$Na provides a beautiful
 example of intruder dominance in the ground states, known experimentally since long
  \cite{Gui84,Mot95}.
 Early  shell model calculations (Poves and Retamosa \cite{Pov87}, Warburton, Becker and Brown \cite{War90})
 unveiled  the role of
 deformed intruder configurations,  2p-2h neutron excitations
 from the $sd$  to the $pf$-shell, and started the study of the boundaries of
 the so called ``island of inversion'' and the properties of its inhabitants.
 Similar mechanisms produce the other known ``islands of inversion''  centered
in   $^{11}$Li  (N=8),  $^{42}$Si  (N=28),  and $^{64}$Cr  (N=40).
We propose now an unified description of the nuclei between Oxygen and Calcium,
covering in many cases all the isotopes between the  neutron and proton drip lines.
The valence space comprises two major shells; 
the $sd$-shell (0d$_{5/2}$, 1s$_{1/2}$,  0d$_{3/2}$)
and the $pf$-shell (0f$_{7/2}$, 1p$_{3/2}$, 1p$_{1/2}$, 0f$_{5/2}$)
and the  effective interaction is SDPF-U \cite{Now09}.

\begin{center}
\begin{figure*}[h]
    \leavevmode
\resizebox{0.6\textwidth}{!}{%
\hspace*{4.5cm}\includegraphics{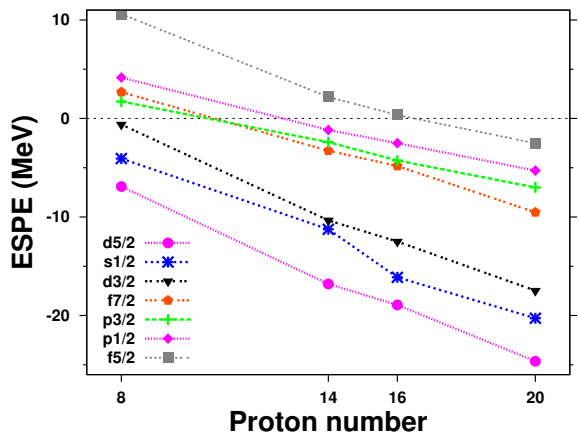}}
  \caption{Neutron effective single particle energies (ESPE) for the N=20 isotones computed with the
  SDPF-U interaction}
  \label{fig:espe}
  \end{figure*}
\end{center}

\vspace*{-1cm}
\medskip
\noindent
N=20: Four protons away from doubly magic $^{40}$Ca,  $^{34}$Si  is 
 a new doubly  magic nucleus because the proton Z=14 and the neutron N=20
 gaps reinforce each other. To go even more neutron rich, one needs to  remove protons from
the  $0d_{5/2}$ orbit. This causes two effects; a reduction of the N=20
neutron gap (see Fig. \ref{fig:espe}) and the increase of proton collectivity.  Both conspire in the sudden
appearance of an Island of Inversion in which Deformed Intruder states become
ground states, as in $^{32}$Mg, $^{31}$Na and $^{30}$Ne.

\medskip
\noindent
N=28: As we remove
 protons from doubly magic $^{48}$Ca, the N=28 neutron gap slowly
 shrinks. In $^{46}$Ar the collectivity induced by the action of the
 four valence protons in the nearly degenerate quasi-spin doublet
 $1s_{1/2}$-$0d_{3/2}$,
 is not enough to beat the N=28
 closure. $^{46}$Ar is non-collective. In $^{44}$S, the quadrupole collectivity 
 sets in. The N=28 closure blows out and prolate and non collective
 states coexist. The ground state and the first excited 2$^{+}$ form
 the germ of a prolate rotational band. In turn $^{42}$Si is an oblate,  well deformed, rotor  
with a first 2$^{+}$ state at 770~keV \cite{Bas07} and  $^{40}$Mg 
is  predicted to be a very collective prolate rotor, with a 2$^{+}$
at $\sim$720~keV.  In addition it could well develop
a neutron halo because more than two
neutrons are, in average, in $p$ wave. 
 
\begin{center}
\begin{figure*}[h]
    \leavevmode
\resizebox{0.5\textwidth}{!}{%
    \includegraphics[angle=0]{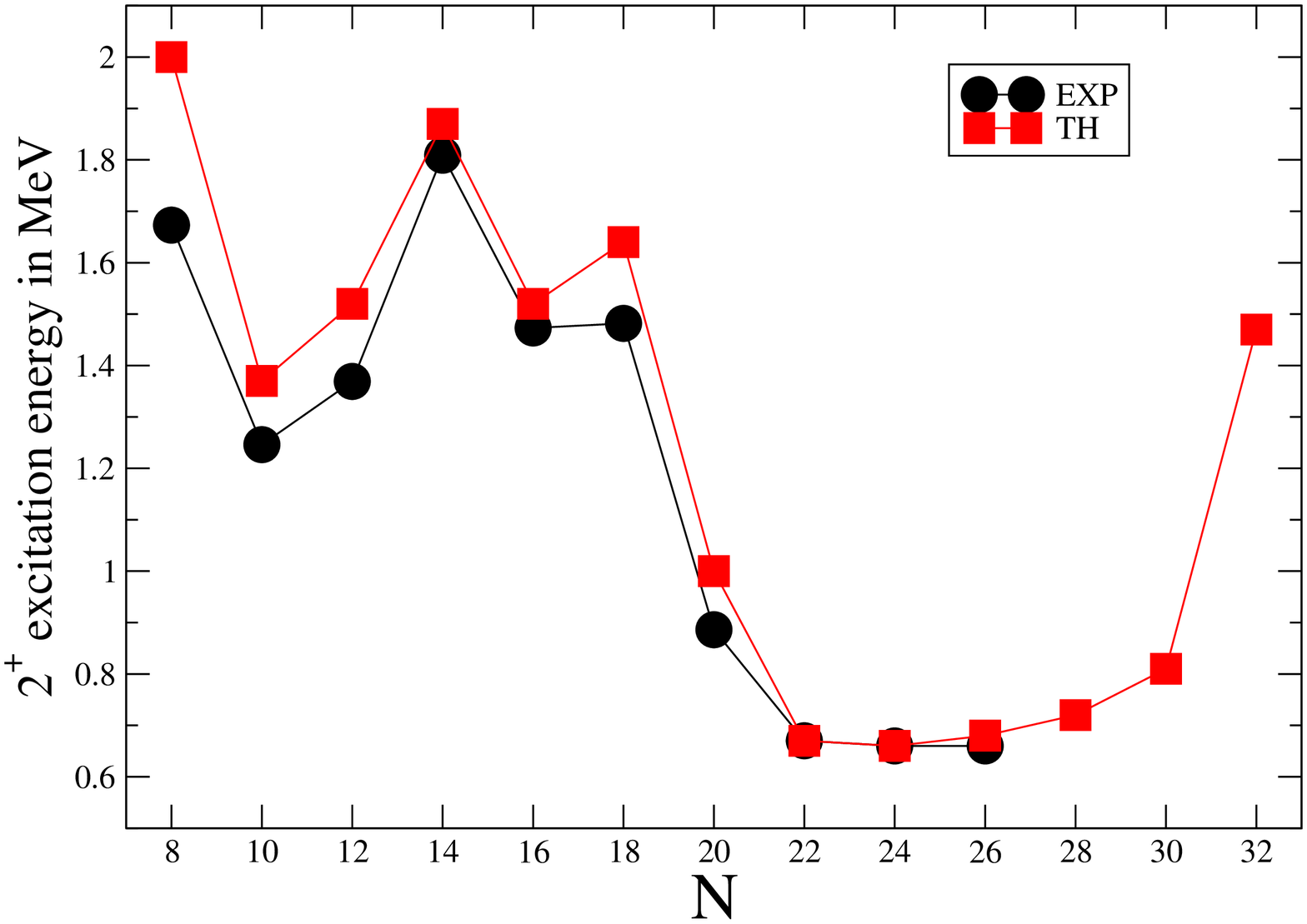}} 
\resizebox{0.5\textwidth}{!}{%
    \includegraphics[angle=0]{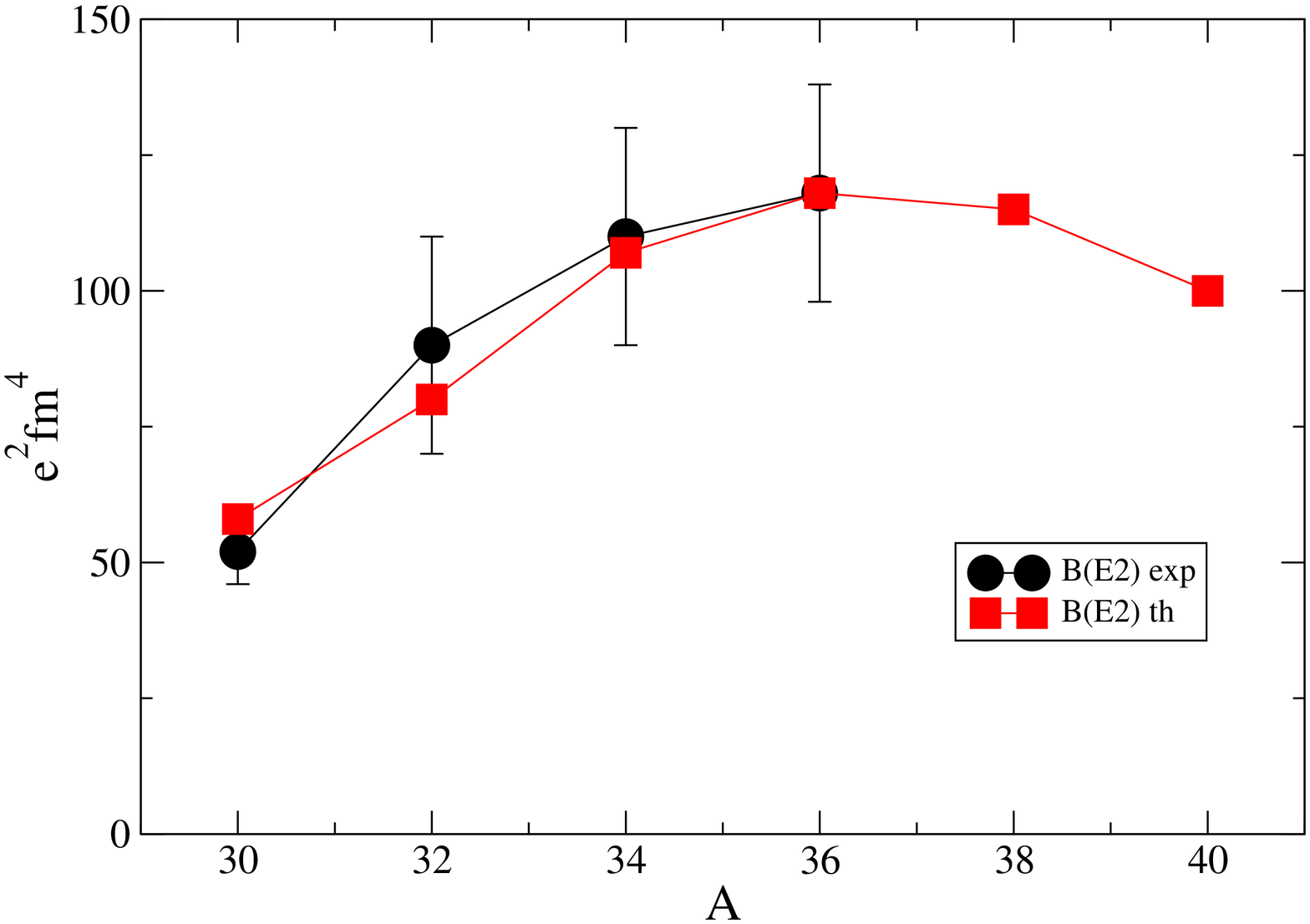}}           
  \caption{Comparison between the theoretical and experimental 2$^+$ excitation energies of the 
  even Mg isotopes (left panel) and B(E2)'s (right panel). In the proton rich side some 
  experimental energies are taken from their mirror nuclei.}
  \label{fig:mg}
  \end{figure*}
\end{center}

 \medskip
 \noindent
 In the left panel of  Fig. \ref{fig:mg} we compare the experimental 2$^+$ excitation energies of the 
  even Mg isotopes with the shell model calculations with the SDPF-U interaction. Up to N=16 the
  calculations are restricted to the $sd$-shell and therefore the results are the same than the ones
  produced by the USD interaction \cite{Bro88}. Beyond N=16 the calculations include up to 6p-6h excitations
  from the $sd$-shell to the full $pf$.  The agreement is excellent and covers all the span of isotopes 
  from the proton to the neutron drip line. Notice the disappearance of the semi-magic closures at
  N=20 and N=28 and the presence of a large region of deformation which connects the two islands
  of inversion, previously though to be split  apart.  In the left panel we compare the B(E2)'s
  in the transition region with some very new experimental data from Riken. The agreement is very good as well.

\vspace*{-1cm} 
 \begin{center}
\begin{figure*}[h]
    \leavevmode
\resizebox{0.5\textwidth}{!}{%
    \includegraphics[angle=0]{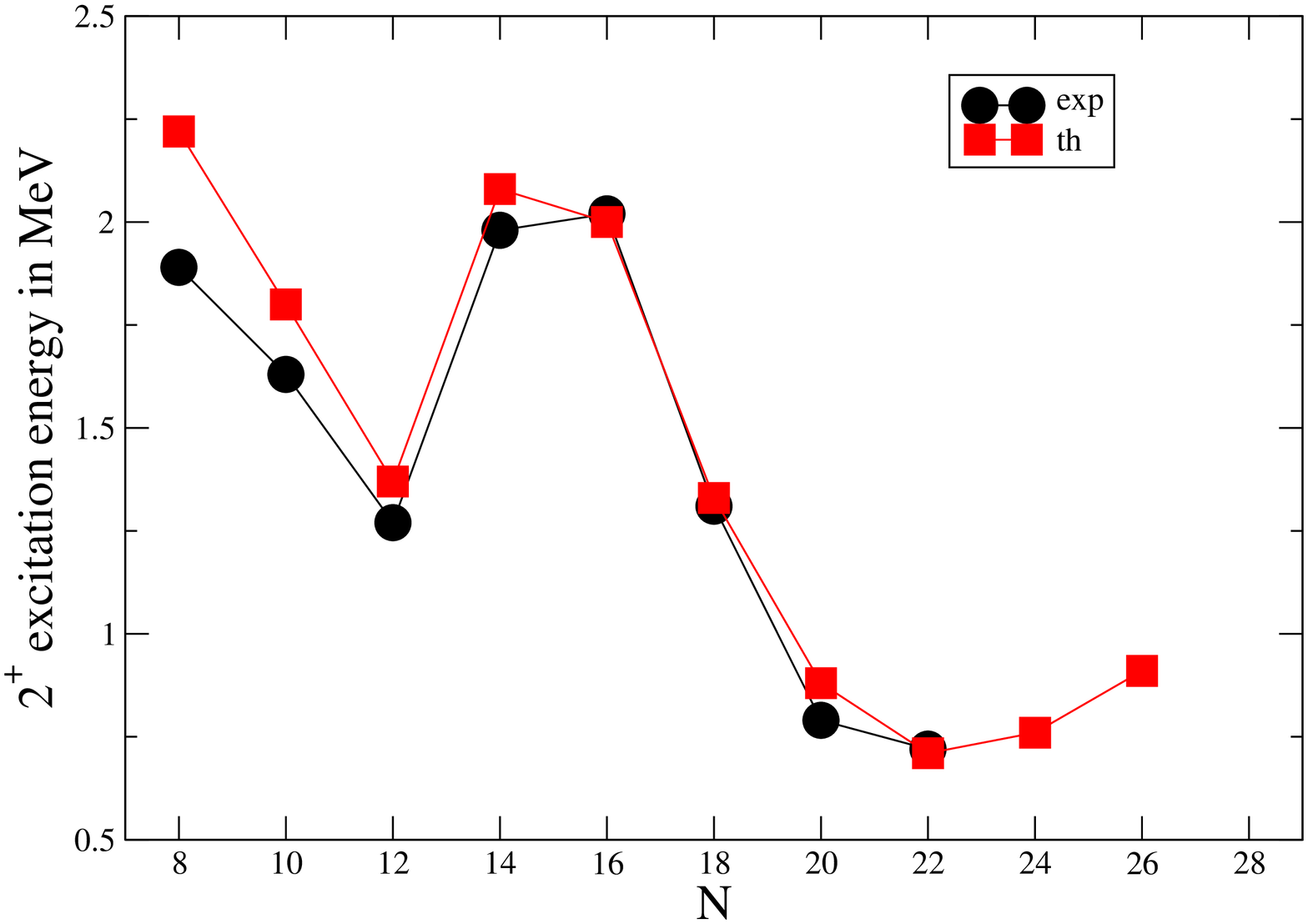}} 
\resizebox{0.5\textwidth}{!}{%
    \includegraphics[angle=0]{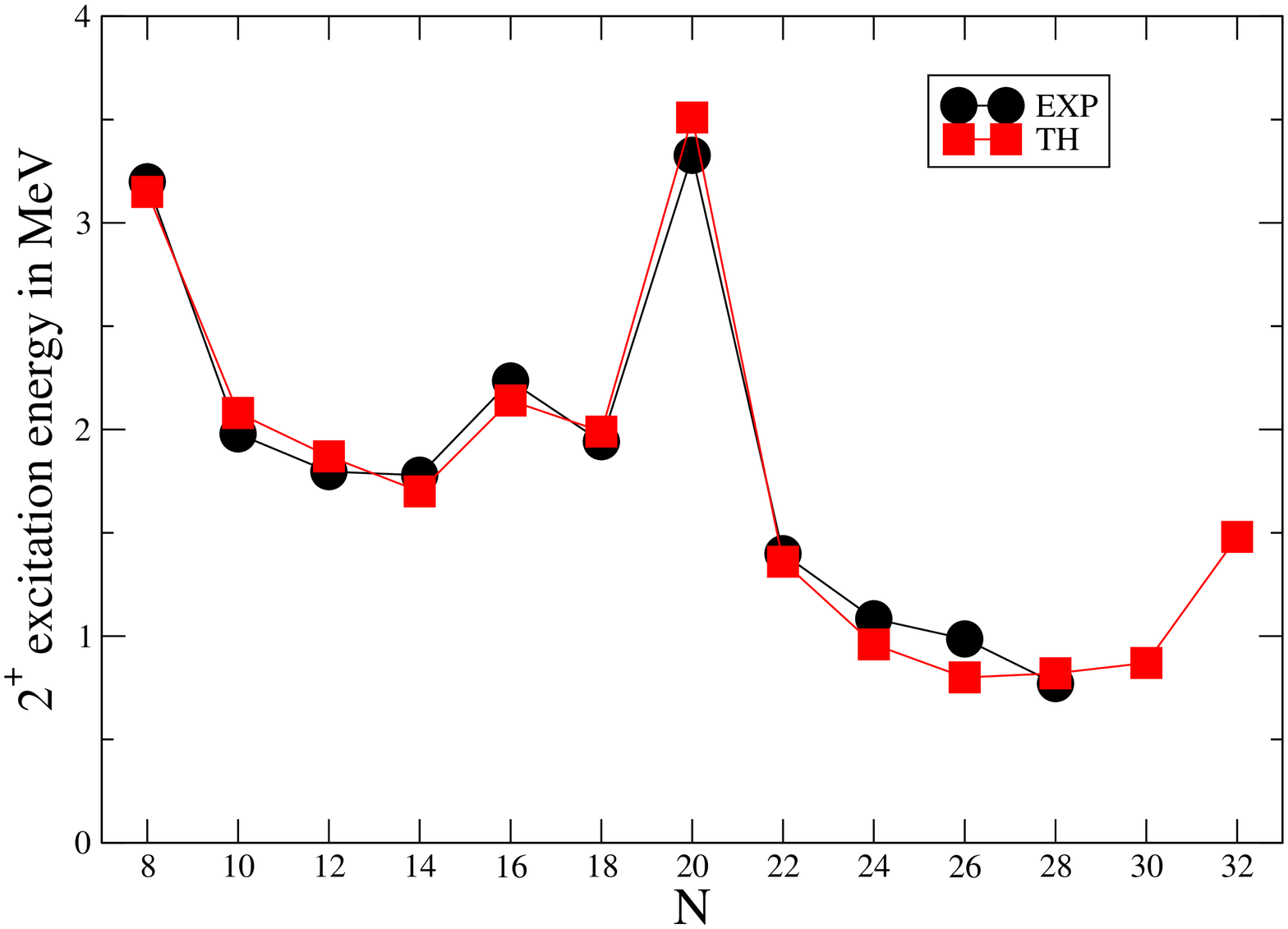}}           
  \caption{Comparison between the theoretical and experimental 2$^+$ excitation energies of the 
  even Ne isotopes (left panel) and the even Si isotopes (right panel). }
  \label{fig:nesi}
  \end{figure*}
\end{center}
 
 \noindent
  The results  for the Neon isotopes  (left panel of Fig. \ref{fig:nesi})  are very similar to the Magnesiums. In the right panel 
  we show the results for the Silicon isotopes  (notice the very different energy scale).   At variance with the Magnesium case, we observe a
  majestic peak at N=20, fingerprint of the double magic nature of $^{34}$Si discussed above and, as in the Ne and Mg cases, no trace
  of the N=28 shell closure is seen.

\section{The island of deformation south of $^{68}$Ni} 
\begin{center}
\begin{figure*}[h]
    \leavevmode
\resizebox{0.5\textwidth}{!}{%
    \includegraphics[angle=0]{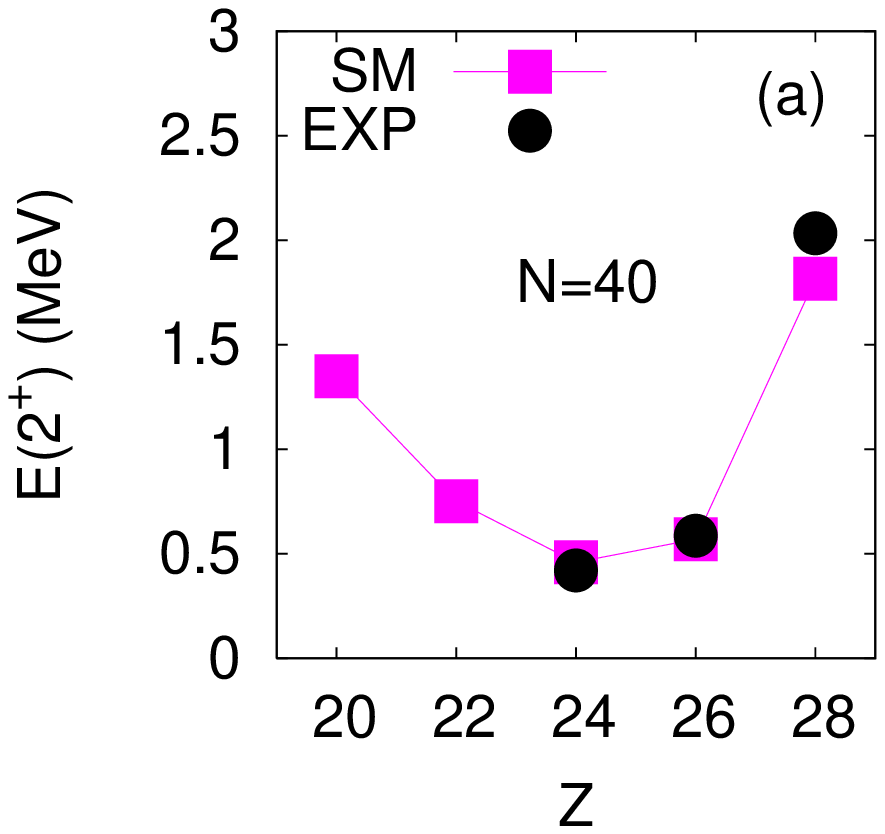}} 
\resizebox{0.5\textwidth}{!}{%
    \includegraphics[angle=0]{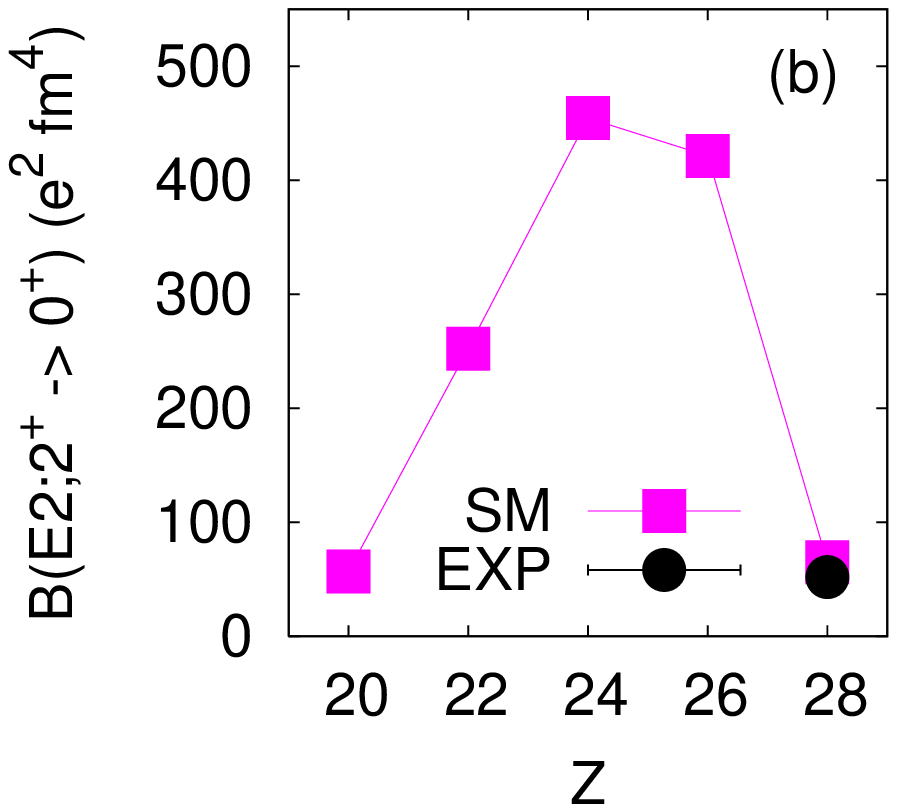}}           
  \caption{The N=40 isotones; comparison between the theoretical and experimental 2$^+$ excitation energies
  (left panel)  and B(E2)'s (right panel) }
  \label{fig:N40}
  \end{figure*}
\end{center}

 The situation at N=40 is similar to the one found at N=20 except
that  $^{68}$Ni is not a ``bona fide''  magic nucleus. 
Removing protons
from the 0f$_{7/2}$ orbit, activates the quadrupole collectivity,
which, in turn, favors the np-nh neutron configurations across
N=40,  which take advantage of the quasi-SU3 coherence of the doublet
 0g$_{9/2}$- 1d$_{5/2}$. Large scale SM calculations in the valence space of the full
 $pf$-shell for the protons and the 0f$_{5/2}$ 1p$_{3/2}$ 1p$_{1/2}$
 0g$_{9/2}$ and 1d$_{5/2}$ orbits for the neutrons, predict a new region of deformation
centered at  $^{64}$Cr.  In Fig. \ref{fig:N40} we show our results for the N=40 isotones: The inversion of configurations sets
 in very rapidly when we remove protons  from $^{68}$Ni, and persists all the way down to $^{60}$Ca
 even in absence of deformation. This shows that the island of 
inversion and the island of deformation may not cover the same territory.  More details on these calculation can be found
in ref. \cite{Len10}.

\vspace*{-1cm}
 \section*{Acknowledgments}
  This work is partly supported by the Spanish Ministry of
 Ciencia e Innovaci\'on under grant FPA2009-13377, by the 
 Comunidad de Madrid (Spain) project HEPHACOS S2009/ESP-1473 and
by the IN2P3(France)-CICyT(Spain) collaboration agreements,

\end{document}